\documentclass{article}
\usepackage[utf8]{inputenc}
\usepackage{graphicx }
\usepackage[figtopcap]{subfigure}
\usepackage[colorlinks,citecolor=blue,urlcolor=blue,linkcolor=blue]{hyperref}
\usepackage{authblk}
\usepackage[sort&compress,numbers]{natbib}
\usepackage{amsmath}	% Advanced maths commands
\usepackage{empheq}
\usepackage{color,soul}

\graphicspath{{./Figs/}}

\title{{Revisiting} Flat Rotation Curves in Chern-Simons Modified Gravity}
\author{Waleed El Hanafy\thanks{waleed.elhanafy@bue.edu.eg}}
\author{Mahmoud Hashim\thanks{mahmoud.hashim@bue.edu.eg}}
\author{G.G.L. Nashed\thanks{nashed@bue.edu.eg}}
\affil{Centre for Theoretical Physics, The British University in Egypt, P.O. Box 43, El Sherouk City, Cairo 11837, Egypt}

\date{ }
\begin{document}

\maketitle
\begin{abstract}
We revisit slow rotating black hole (BH) solutions in Chern-Simons modified gravity (CSMG) by considering perturbative solution about Schwarzschild BH. In particular, the case when nondynamical CSMG with noncanonical CS scalar is considered. We provide a new solution different from the previously obtained one \cite{Konno:2007ze} which we refer to as KMT model. The present solution accounts for frame dragging effect which includes not only radial dependence as in the KMT. Nevertheless, it reduces to KMT as a particular case. We show that the tidal gravitational force (Kretschmann invariant) associated to the present solution contains a term of order $1/r^3$ additional to Schwarzschild but absent from directional divergence, unlike KMT model which diverges along the axis of symmetry. We derive the corresponding circular velocity of a massive test particle in which the KMT velocity is recovered in addition to an extra term $\propto r$. We investigate possible constraints on KMT and the present solutions from the observed rotation curve of UGC11455 galaxy as an example. We show that perturbation solutions cannot physically explain the flattening of galactic rotation curves.
\end{abstract}
%%%%%%%%%%%%%%%%%%%%%%%%%%%%%%%%%%%%%%%%%%%%%%%%%%%%%%%%%%%%%%%%%%%%%%%
\section{Introduction}
\label{sec:introduction}
%%%%%%%%%%%%%%%%%%%%%%%%%%%%%%%%%%%%%%%%%%%%%%%%%%%%%%%%%%%%%%%%%%%%%%%%
The general relativity (GR) theory is no doubt setting the solid bases which formulate the contemporary approaches to understand nature of gravity. The astonishing idea that gravity is a curvature of pseudo-Riemann spacetime has received recognition over the years by passing many observational tests like  planetary motion in the solar system \cite{RevModPhys.19.361}, bending of light \cite{doi:10.1098/rsta.1920.0009}, binary pulsars \cite{1981SciAm.245d..74W,2010ApJ...722.1030W} (see also \cite{Will:2014kxa}), gravitational waves detection due to binary compact objects merging by Laser Interferometer Gravitational-Wave Observatory (LIGO) and Virgo Collaborations \cite{LIGOScientific:2016aoc,LIGOScientific:2017ycc,LIGOScientific:2017vwq} and recently BH shadows by Event Horizon Telescope \cite{EventHorizonTelescope:2019dse}. However, it struggles with inconsistent description of gravity on the scales of galaxies. The observed velocities at large radii from the galactic centres are inconsistent with the gravitational field from the visible light from stars only, which are mostly clustered in the middle of galaxies. Unlike the solar system planetary motion, the velocities of the stars away from the centre of a galaxy have almost same velocities of the stars near the core, which results in a flattening rotational velocity curve at large distances within galaxies \citep[c.f.][]{1970ApJ...159..379R,1980ApJ...238..471R,1991MNRAS.249..523B,Corbelli:1999af}.

One way to explain the stellar motion is to assume that the galaxies are embedded in large Cold Dark Matter (CDM) haloes as noted early by Zwicky in 1933. The temperature anisotropy of the cosmic microwave background (CMB) spectrum as detected by Planck provides a strong evidence supporting existence of CDM \citep{Planck:2018vyg}. Weakly interacting massive particles (WIMPs) within an energy scale GeV--TeV have been suggested as candidates to represent CDM from particle physics perspective. The hypothesised WIMPs perfectly match the CDM abundance observed today \citep{Jungman:1995df}. However, still no experimental evidences support this hypothesis, in fact, it is mostly excluded by large hadron collider (LHC) \citep{Craig:2013cxa}, or other detectors \citep[c.f.,][]{COSINE-100:2021xqn}. In this sense, recent experiments significantly narrow possibilities to explain CDM as particles.

Another way to explain the flat rotation curve is to investigate possible modification of gravity at these scales. This could be reasonable; since the assumed CDM could not mediate electromagnetic force and it can be seen only through its gravitational effect on stellar motion. It has been suggested that the Newtonian acceleration law at weak gravity regimes should be modified as noted by \cite{1983ApJ...270..365M,1983ApJ...270..371M} in Modified Newtonian Dynamics (MOND). Within this frame, it is possible to gain some successes, e.g. the explanation for the flat galactic rotation curves and the Tully-Fisher relation without the need for CDM particles. Notably, the longstanding struggle of MOND explaining CMB power spectra has been reinvestigated for a new relativistic MOND theory, with some phenomenological requirements, the recent version of the theory agrees with the observed CMB and matter power spectra on linear cosmological scales \citep{Skordis:2020eui}. On the other hand, a sample of 152 galaxy rotation curves included by Spitzer Photometry and Accurate Rotation Curves (SPARC) database, has been used to distinguish CDM halos, modified gravity and modified inertia \citep{Chae:2022oft}. It has been shown that CDM halos predict a systematically deviating relation from the observed one. All aspects of rotation curves are most naturally explained by modified gravity. Several modified gravity theories have been introduced to explain the flattening of the rotational velocity curves of spiral galaxies \citep[c.f.,][]{moffat2004modified, Bertolami:2007gv, Harko:2011kv,  Moffat:2013sja,Capozziello:2013yha, Moffat:2014pia, Lelli:2016zqa, Panpanich:2018cxo, Chae:2020omu, Chae:2022oft, Burikham:2023bil}. Remarkably, one could relate BH solutions to rotation curves of galaxies following a simple approach: Solve the field equation of the suggested gravity theory to obtain BH vacuum solution. Obtain the first integral of Euler-Lagrange equation and deduce the circular motion $v_{circ}$ of a test particle of mass $m$. In the case of galaxies, the space is not exactly vacuum, it rather has visible baryons with mass distribution $M(r)$, that is similar to stellar models. However, the visible matter pressure in the galaxy is very small, the relativistic effects of the pressure of the baryonic matter are negligible. Therefore, the setup is consistent with Tolman-Oppenheimer-Volkof (TOV) equation at weak field limit, whereas $M(r)$ for each galaxy is governed by
\mbox{$M(r)/r=V^2_\text{bar}=\sum_i \Upsilon_i V_i^2$,}
where $i$ denotes the bulge, disk and gas components. Thus, one uses observational data of the velocity contributions $V_i$ for each component to fit the parameter space including the coefficients $\Upsilon_i$ and additional parameters due to modified gravity. It remains unclear how modified gravity can explain Bullet Cluster with no dark matter \citep{Clowe:2003tk,Markevitch:2003at}, see also \citep{Pardo:2020epc}.

CSMG is one of the well inspired modifications of GR, since it is motivated by anomaly cancellation in particle physics, string theory and loop quantum gravity. In particle physics, motivated by the CS modification of electrodynamics \citep{Carroll:1989vb}, gravitational ABJ anomaly can be cancelled by adding CS term into Einstein-Hilbert action \citep{Jackiw:2003pm}. In string theory, the coupling of a gauge field to the string results in a gravitational anomaly in the effective low energy 4-dimensional GR \citep{ALVAREZGAUME1984269}, it has been shown that a CS term should arise via Greens-Schwarz anomaly cancelling mechanism \citep{2012ssti.book.....G}. In the effort to quantise GR within the loop quantum gravity framework by considering discrete spacetime, the investigation of parity and charge-parity conservation leads to CS modification \citep{1989IJMPA...4.1493A}. For more details see the review \citep{Alexander:2009tp}.

In general, there are two distinct settings in which the CSMG has been formulated, those are non-dynamical and dynamical CSMG. The pseudo CS scalar field is understood as a wholly external quantity in the former case, while it has dynamics in the latter fulfilling its own equation of motion. Although Kerr metric is not a solution in CSMG \citep[c.f.,][]{Grumiller:2007rv,Alexander:2009tp}, spinning BH solutions can be found. We may classify the existing solutions as follows: (i) Exact solutions using a non-dynamical CS scalar, it has been shown that, for the standard choice of the CS scalar, the field equations eliminate all solutions except Schwarzschild \cite{Grumiller:2007rv}. On the other hand, for a noncanonical choice of the CS scalar, \textit{mathematical} rotating BH solutions can be found. (ii) Exact solutions using a dynamical CS scalar have not been found yet as far as we know. (iii) The slow-rotating approximation using a non-dynamical CS scalar has been studied in \cite{Konno:2008np} for a noncanonical CS scalar, where the circular velocity reads $v_{circ} \sim \sqrt{M/r} + \text{const}$ far from the galactic centre. This solution has been proposed to explain the flattening of the rotation curve due to the existence of the additional constant. In the same line, a generalised choice of the scalar field has been suggested \cite{Yunes:2009hc}, where the circular velocity $v_{circ} \sim \sqrt{M/r} - M/r^2$, which cannot explain the galactic rotation curve without dark matter. (iv) The slow-rotating approximation using a dynamical CS scalar has been studied in \cite{Yunes:2009hc}, which shows that the full dynamical solution introduces extra terms that are highly suppressed by large inverse powers of the radius. Consequently, it would not be able to describe the flattening of the rotational velocity curve. On the other hand, CSMG modifies early universe leaving testable features on the observed CMB polarisation on the superhorizon scales at low multipole moments \cite[c.f.,][]{Lue:1998mq,Alexander:2006mt}. Notably, these corrections are practically hidden due to large cosmic variance at these scales.

In the present work, we revisit slow rotating BHs solution in CSMG using the perturbation expansion around the Schwarzschild. In particular the nondynamical CSMG with noncanonical CS scalar which has been proposed by {\cite{Konno:2007ze}} and will be referred to as KMT model in the present study. As just mentioned above, this slow-rotating BH solution has been advocated to explain the flatness of rotational velocity curves of spiral galaxies {\citep{Konno:2008np}}. However, it never be confronted with observational data. We organise the paper as follows: In Sec. \ref{sec:CS-gravity}, we briefly review the nondynamical CS modification of GR gravity. We further apply the field equations to stationary and axially symmetric spacetime configuration which represents gravitational field of slowly rotating BH. We obtain a new CS solution which reduces the previously obtained solution by KMT as a special case. We discuss the physical features of the new solution in comparison to the known Kerr solution and KMT model by comparing Kretschmann invariant in each case. In Sec. \ref{sec:rotation-curve}, we determine the geodesic motion of a massive test particle in the gravitational field of the obtained CS solution. Then, we obtain the CS rotational velocity of the circular and equatorial motion which contains an additional term, proportional to the radial distance, in comparison to KMT model. In Sec. \ref{sec:data_model}, we use the rotation curve of UGC11455 galaxy to investigate possible constraints on the parameter space of the KMT and the present model. This shows clearly the failure of KMT model to fit the curve in principle. Although the present model shows good results, it cannot fulfil the perturbation approach assumption to interpret the rotation curve. This conclusion should be valid for any perturbative solution around Schwarzschild BH. In Sec. \ref{sec:Summary}, we summarise and conclude the obtained results.
%%%%%%%%%%%%%%%%%%%%%%%%%%%%%%%%%%%%%%%%%%%%%%%%%%%%%%%%%%%%%%%%%%%%%%%
\section{Slowly Rotating Black Hole in General Relativity and Chern-Simons Gravity}
\label{sec:CS-gravity}
%%%%%%%%%%%%%%%%%%%%%%%%%%%%%%%%%%%%%%%%%%%%%%%%%%%%%%%%%%%%%%%%%%%%%%%
We give a brief review on Chern-Simons modifications of the GR theory. In particular, nondynamical CSMG with noncanonical CS scalar as discussed by \cite{Konno:2007ze}. For more general framework with dynamical CSMG, see the review \citep{Alexander:2009tp}. Unless otherwise noted, we use the geometric units\footnote{The geometric units will be used in Sections \ref{sec:CS-gravity}--\ref{sec:rotation-curve} to write the equations in simplified forms, while the standard physical units will be restored in {Section {\ref{sec:data_model}} for practical purposes}.} in which the gravitational constant and the speed of light are set to $G=c=1$, whereas the action can be written as
\begin{equation}
    I=\frac{1}{16\pi} \int d^4 x \left(-\sqrt{-g}R+\frac{1}{4} l \, \vartheta \,\, ^\ast{R}{^\sigma}{_\tau}{^{\mu\nu}} R{^\tau}_{\sigma\mu\nu}\right).
\end{equation}
The first term gives Einstein\textendash{Hilbert} action with metric determinant $g$ and Ricci scalar $R$. The second term gives CS modifications with a coupling constant $l$, an external quantity $\vartheta$ acts as a coupling field in CS modified gravity and dual Riemann tensor $^\ast{R}{^\sigma}{_\tau}{^{\mu\nu}}=\frac{1}{2} \varepsilon^{\mu\nu\alpha\beta} R^{\tau}{_{\sigma\alpha\beta}}$ where $\varepsilon$ denotes Levi-Civita tensor. For constant $\vartheta$, the GR theory is reproduced as a particular case \citep{Alexander:2009tp}.

The variation with respect to the metric tensor $g_{\mu\nu}$ gives rise to the following field equations
\begin{eqnarray}
G^{\mu\nu}+l C^{\mu\nu}&=&-8\pi T^{\mu\nu}, \label{eq:F-eq1}\\
    0=\nabla_\mu C^{\mu\nu}&=&\frac{1}{8\sqrt{-g}} v^\nu \, \ ^\ast{R}{^\sigma}{_\tau}{^{\mu\lambda}} R{^\tau}_{\sigma\mu\lambda}, \label{eq:F-eq2}
\end{eqnarray}
where $G^{\mu\nu}$ is Einstein tensor, $T^{\mu\nu}$ is the stress-energy tensor, $v_\alpha=\partial_\alpha \vartheta$ is called embedding vector and $C^{\mu\nu}$ is Cotton tensor,
\begin{eqnarray}\label{eq:Cotton}
    C^{\mu\nu}&=&-\frac{1}{2\sqrt{-g}}\Bigg[\Bigg. v_\sigma \left(\varepsilon^{\sigma\mu\alpha\beta}\nabla_\alpha R{^\nu}{_\beta} + \varepsilon^{\sigma\nu\alpha\beta}\nabla_\alpha R{^\mu}{_\beta}\right)+v_{\tau\sigma} \left(^\ast{R}^{\tau\mu\sigma\nu}+^\ast{R}^{\tau\nu\sigma\mu}\right)\Bigg.\Bigg],
\end{eqnarray}
with $v_{\tau\sigma}=\nabla_{\sigma}v_\tau$ and $\nabla$ is the covariant derivative associated to Levi-Civita linear connection. By Bianchi identity $\nabla_\mu G^{\mu\nu}\equiv 0$ and by requiring the matter conservation $\nabla_\mu T^{\mu\nu}=0$, the free-divergence Cotton tensor constraint should be imposed, namely Equation \eqref{eq:F-eq2}. For a solution to fulfil CS gravity, it should satisfy Equations \eqref{eq:F-eq1} and \eqref{eq:F-eq2}. Therefore, those are the basic equations which govern CSMG in the present context.

It is worth mentioning that for a spherically symmetric spacetime the CS solution is nothing but Schwarzschild solution, since that the Cotton tensor vanishes identically. For a slowly rotating BH, we consider perturbation around the Schwarzschild solution, where the expansion parameter $\epsilon$ is related to the BH angular momentum $J$, i.e. $J \sim O (\epsilon)$. Thus, the spacetime configuration gives rise to a stationary and axisymmetric metric in a spherical polar coordinates ($t$, $r$, $\theta$, $\phi$) \citep{Konno:2007ze,Konno:2008np}
\begin{equation}\label{eq:slow-rot-sptime}
    ds^2=-A(r) dt^2 +\frac{dr^2}{A(r)} + r^2(d\theta^2+ \sin^2{\theta}\, d\phi^2)-2 r^2 \omega(r,\theta) d\phi dt,
\end{equation}
where the time coordinates $-\infty<t<\infty$, the radial distance $0\leq r<\infty$, the zenith (inclination) angle $0\leq \theta\leq \pi$, the azimuthal angle $-\pi\leq \phi<\pi$, the metric potential $A(r)\equiv 1-\frac{2M}{r}$ and $\omega(r,\theta)$ represents dragging of local inertial frames, i.e. Lense-Thirring effect. Independent of the choice embedding coordinate, it has been shown that no rotating BH solution can be found up to the first order whereas the embedding coordinate is timelike, i.e. $\vartheta \propto t$, unless the limit $M/r \to 0$ is fulfilled \citep{Konno:2007ze}. On the contrary, for the choice $\vartheta \propto r \cos \theta$, i.e.  the embedding coordinate is spacelike, rotating BH solution can be found for arbitrary BH mass. In the following subsections, we review a previously obtained solution limited to a particular choice for the frame-dragging, then we obtain the generalised solution with a comparison to the GR solutions in each case.

\subsection{Case I: KMT model}
\label{Sec:CaseI}

For a spacelike embedding coordinate $\vartheta = r \cos \theta/\lambda_0$, the embedding vector is written as $v_\alpha$ = ($0$, $\cos \theta/\lambda_0$, $-r \sin \theta/\lambda_0$, $0$) \citep{Konno:2007ze}. Applying the free-divergence Cotton tensor \eqref{eq:F-eq2} to the spacetime \eqref{eq:slow-rot-sptime}, we obtain the following constraint up to $O(\epsilon)$
\begin{equation}\label{eq:Cott-div-free}
    \omega_{,r\theta}=0,
\end{equation}
where the sub-indices $,r$ and $,\theta$ denote the derivatives with respect to those coordinates. It has been shown that for the particular choice $\omega(r,\theta)\to \omega(r)$ the above constraint is satisfied. Additionally, it can be shown that the field equations \eqref{eq:F-eq1} reduce to one equation
\begin{equation}\label{eq:CS-omegar}
    r(r-2M)\omega_{,rr}+4(r-2M)\omega_{,r}+2\omega=0.
\end{equation}
Therefore, the solution has been given as \cite{Konno:2007ze,Konno:2008np}
\begin{equation}\label{eq:CS-omegar-soln}
    \omega(r)=\frac{C_1}{r^2}\left(1-\frac{2M}{r}\right)+\frac{C_2}{r^3}\left[r^2-2Mr-4M^2+4M(r-2M)\ln(r-2M)\right],
\end{equation}
where the constants $C_1$ and $C_2$ are of order $O(\epsilon)$ whereas $\epsilon \equiv J/Mr$ with a total angular momentum $J$. In fact, the choice of the drag effect to be radial $\omega(r,\theta)\to \omega(r)$ satisfies $\omega_{,r\theta}\equiv 0$ identically. It, additionally, makes the Cotton tensor components to vanish up to $O(\epsilon)$. Thus, the field equation \eqref{eq:CS-omegar} is identical to the GR field equation $G^{t\phi}=0$, whereas the other components vanish identically. In this sense, we conclude that the particular choice of the frame drag to have a radial dependence only hides the CS gravity contribution rendering it equivalent to GR gravity up to $O(\epsilon)$. This has been already noted by \cite{Konno:2007ze,Konno:2008np}, we refer to the solution \eqref{eq:CS-omegar-soln} as KMT model. In the following subsection, we drop this constraint seeking for a new CS gravity solution distinguished from the GR one.

\subsection{Case II: CSMG model}
\label{Sec:CaseII}
For a spacelike embedding coordinate $\vartheta = r \cos \theta/\lambda_0$, we write the embedding vector $v_\alpha = (0,\cos \theta/\lambda_0,-r \sin \theta/\lambda_0,0)$. Applying the field equations \eqref{eq:F-eq1} to the spacetime \eqref{eq:slow-rot-sptime}, up to the order $O(\epsilon)$, we obtain the non-vanishing components ($t\phi$), ($rr$), ($r\theta$), ($\theta\theta$), ($\phi\phi$) and ($tt$), respectively, as follows
\begin{eqnarray}
    &&r(r-2M)\omega_{,rr}+\omega_{,\theta\theta}+4(r-2M)\omega_{,r}-\cot\theta\, \omega_{,\theta}+2\omega = 0,\qquad \label{eq:t-phi}\\ [5pt]
    &&r(r^2-5rM+6M^2)\omega_{,rr}+(r-3M)\omega_{,\theta\theta}-3M(r-2M) \cot\theta \omega_{,r\theta}+(r-3M)\left[4(r-2M)\omega_{,r}-\cot\theta \omega_{,\theta}+2\omega\right]= 0,\qquad \label{eq:r-r}\\ [5pt]
    &&(r-3M)\cot\theta \omega_{,\theta\theta}+r^2(r-2M)\cot\theta\omega_{,rr}+\left[(r-6M)-\sec^2\theta (r-3M)\right]\omega_{,\theta} + 2r\cot\theta \left[2(r-2M)\omega_{,r}+\omega\right]= 0,\qquad \label{eq:r-theta}\\ [5pt]
    &&r(r-2M)\omega_{,rr}+\omega_{,\theta\theta}+3M\tan\theta\omega_{,r\theta}+4(r-2M)\omega_{,r}-\cot\theta\omega_{,\theta}+2\omega = 0, \qquad \label{eq:theta-theta}\\ [5pt]
    &&r^2 \tan\theta (r-2M)\omega_{,rrr}+\omega_{,\theta\theta\theta}+r(r-2M)\omega_{,rr\theta}+r \tan\theta \omega_{,r\theta\theta}-r\tan\theta\left[(r-2M)\sec^2 \theta-2(3r-5M)\right]\omega_{,rr}\nonumber \\
    &&-\frac{1+\cos^2 \theta}{\sin\theta \cos\theta}\omega_{,\theta\theta}+(3r-5M)\omega_{,r\theta}-2\tan\theta\left[2(r-2M)\sec^2 \theta-3r\right]\omega_{,r}+2\frac{1+\sin^2 \theta}{\sin^2 \theta}\omega_{,\theta}-2\sec\theta \csc\theta \omega=0, \qquad \label{eq:phi-phi}\\ [5pt]
    &&r^2 \sin\theta \left[4M^2+r(r-4M)\right]\omega_{,rrr}+\cos\theta (r-2M)\omega_{,\theta\theta\theta}+r\cos\theta \left[4M^2+r(r-4M)\right]\omega_{,rr\theta}+r\sin\theta (r-2M) \omega_{,r\theta\theta}\nonumber\\
    &&+r\sin\theta(6r^2-23 r M +22 M^2)\omega_{,rr}+\left[\sin\theta(r-3M)-\sec\theta(r-2M)\right]\omega_{,\theta\theta}+\cos\theta(3r^2-11 r M +10 M^2)\omega_{,r\theta}\nonumber\\
    &&+2\sin\theta (3r^2-8 r M+ 4M^2)\omega_{,r}+\cos\theta \left[(2r-3M)+\sec^2\theta(r-2M)\right]\omega_{,\theta}+2M\cos\theta \cot\theta \omega= 0. \qquad \label{eq:t-t}
\end{eqnarray}

In addition, the free-divergence Cotton tensor, namely equation \eqref{eq:F-eq2}, up to $O(\epsilon)$, gives rise to
\begin{equation}\label{eq:Cott-div-free2}
    \omega_{,r\theta}=0.
\end{equation}
On the contrary to the previous solution (KMT) \citep{Konno:2007ze}, we take the general solution of the above differential equation $\omega(r,\theta)=\omega_{r}(r)+\omega_{\theta}(\theta)$. Then, we find the solution of the system, namely Equations \eqref{eq:t-phi}--\eqref{eq:t-t}, as follows
\begin{eqnarray}\label{eq:CS-omega-soln}
\nonumber    \omega(r,\theta)&=&\frac{C_1}{r^2}\left(1-\frac{2M}{r}\right)+\frac{C_2}{r^3}\left[r^2-2Mr-4M^2+4M(r-2M)\ln(r-2M)\right]\\
    &+&\frac{C_3}{r^3}\Bigg\{ r^3(1-\cos 2\theta )-6M\left[5M^2-(r-3M)^2\right]+24M^2(r-2M)\ln(r-2M)  \Bigg\},
\end{eqnarray}
Clearly, the above solution differs from the one given in Equation \eqref{eq:CS-omegar-soln} by an additional term which contains the constant $C_3$. The obtained solution, Equation \eqref{eq:CS-omega-soln}, is a new one which generalises the previously obtained solution \cite{Konno:2007ze}, i.e. Equation \eqref{eq:CS-omegar-soln}. Consequently, we evaluate the cross term of the metric component
\begin{eqnarray}
    \nonumber    g_{t\phi}&=& C_1\left(1-\frac{2M}{r}\right)+\frac{C_2}{r}\left[r^2-2Mr-4M^2+4M(r-2M)\ln(r-2M)\right]\\
      &+&\frac{C_3}{r}\Bigg\{ r^3(1-\cos 2\theta )-6M\left[ 5M^2-(r-3M)^2\right]+24M^2(r-2M)\ln(r-2M)  \Bigg\}.
\end{eqnarray}
At the limit $r \gg M$, we have
\begin{equation}
    g_{t\phi}=C_1+C_2 r + 2 C_3 r^2 \sin^2 \theta.
\end{equation}
The above equation indicates that the frame-dragging effect works in the whole space as some terms diverge as $r$ increases.

In the previous subsection, we clarified that the CS and GR provide the same solution once the frame dragging effect is chosen to be radial function $\omega(r)$. In order to check the novelty of the just obtained solution by accounting for CS terms in comparison to GR, assuming the general case that the dragging effect is $\omega(r,\theta)$, we write the GR field equation up to $O(\epsilon)$ as follows
\begin{equation}\label{eq:Gtphi}
    (r-2M)r\omega_{,r\theta}+\omega_{,r\theta}+4(r-2M)\omega_{,r}-\cot \theta \omega_{,\theta}+2\omega(r,\theta)=0.
\end{equation}
The above equation gives the component $G^{t\phi}=0$, whereas other components vanish identically. This leads to the solution
\begin{eqnarray}\label{eq:GR-omega-soln}
    \omega(r,\theta)&=&\sin^2 \theta \left[\tilde{C}_1 (r-2M)^{\frac{-3+\alpha}{2}} F_1
    + \tilde{C}_2 (r-2M)^{\frac{-3-\alpha}{2}} F_2 \right]\left[\tilde{C}_3 F_3 + \tilde{C}_4 F_4 \cos \theta \right].
\end{eqnarray}
where $\tilde{C}_i$ are integration constants and
\begin{eqnarray*}
    F_1&=&\text{hypergeom}\left(\left[{\frac{3-\alpha}{2}},{\frac{5-\alpha}{2}}\right],[1-\alpha],-\frac{2M}{r-2M}\right),\\
    F_2&=&\text{hypergeom}\left(\left[{\frac{3+\alpha}{2}},{\frac{5+\alpha}{2}}\right],[1+\alpha],-\frac{2M}{r-2M}\right),\\
    F_3&=&\text{hypergeom}\left(\left[{\frac{3+\alpha}{4}},{\frac{3-\alpha}{4}}\right],\left[\frac{1}{2}\right],\cos^2 \theta\right),\\
    F_4&=&\text{hypergeom}\left(\left[{\frac{5+\alpha}{4}},{\frac{5-\alpha}{4}}\right],\left[\frac{3}{2}\right],\cos^2 \theta\right),\\
    \alpha&=&\sqrt{9-4 c_1},\,c_1\text{is an arbitrary constant}.
\end{eqnarray*}
Obviously, the GR field equation \eqref{eq:Gtphi} coincides with one of the CS field equations, the ($t\phi$)-component, namely Equation \eqref{eq:t-phi}. Consequently, the CS solution \eqref{eq:CS-omega-soln} is a GR solution as well. On the contrary, the GR solution \eqref{eq:GR-omega-soln} does not fulfil CS field equations \eqref{eq:r-r}--\eqref{eq:Cott-div-free2}. Therefore, we ignore the GR solution of rotating BH since the exact Kerr BH solution is well known.

We conclude that solution \eqref{eq:CS-omega-soln} provides a new solution which manifests novel features of CS gravity, while it still satisfies the GR gravity. The conclusion is not true for the opposite direction. In the following section, we discuss some physical features of interest for the obtained solution in comparison to some other rotating BH solutions.
%%%%%%%%%%%%%%%%%%%%%%%%%%%%%%%%%%%%%%%%%%%%%%%%%%%%%%%%%%%%%%%%%%%%%%%
\subsection{Physical Features of Solution}
\label{sec:Phys}
%%%%%%%%%%%%%%%%%%%%%%%%%%%%%%%%%%%%%%%%%%%%%%%%%%%%%%%%%%%%%%%%%%%%%%%
In practice, it is always useful to quantify the amount of curvature as a function of position near the BH to examine the genuine tidal force of gravity for any solution. We use a quadratic scalar invariant, that is Kretschmann scalar $R^{\alpha \beta \gamma \delta} R_{\alpha \beta \gamma \delta}$ for that purpose, and also to compare different solutions. We begin with the well known Kerr metric, with $g^{(K)}_{t\phi}=-2 J_0 \sin^2 \theta / r$ where $J_0\sim O(\epsilon)$, then
\begin{equation}\label{eq:Ksq-Kerr}
    R_{(K)}^{\alpha \beta \gamma \delta} R_{(K) \alpha \beta \gamma \delta} \simeq \frac{48 M^2}{r^6} - \frac{144 J_0^2}{r^8}(2+\cos 2\theta).
\end{equation}
The first term is identical to Schwarzschild BH while the second term is originated by the frame-dragging effect. Clearly, the latter is of order $r^{-8}$, which decays more rapidly than the Schwarzschild part.

For the slowly rotating BH solution \eqref{eq:CS-omegar-soln}, assuming that $\omega(r,\theta) \to \omega(r)$, we calculate Kretschmann invariant up to $O(\epsilon)$
\begin{eqnarray}\label{eq:Ksq-omegar}
\nonumber     R^{\alpha \beta \gamma \delta} R_{\alpha \beta \gamma \delta} &\simeq& \left(R^{(0)\alpha \beta \gamma \delta}+R^{(1)\alpha \beta \gamma \delta} \right) \left(R^{(0)}{_{\alpha \beta \gamma \delta}} + R^{(1)}{_{\alpha \beta \gamma \delta}} \right)\\
     &\simeq& \frac{48 M^2}{r^6} - \frac{4 C_2^2}{r^4 \sin^4 \theta},
\end{eqnarray}
where $R^{(0)\alpha \beta \gamma \delta}$ refers to the zeroth order $\epsilon$-quantity constructed from Schwarzschild spacetime and $R^{(1)\alpha \beta \gamma \delta}$
 refers to the first order $\epsilon$-quantity constructed from the frame-dragging effect. This is in agreement with the obtained value in \cite{Konno:2008np}. In comparison to Kerr solution the first term is common and originated by Schwarzschild metric, while the second term, $O(r^{-4})$, is due to rotation which decays more slowly than the corresponding term in Kerr and than Schwarzschild term as well. The solution is singular along the axis of rotation ($\theta=0$, $\pi$), it is asymptotically flat otherwise.

In the present study, dropping the frame-dragging assumption $\omega(r,\theta) \to \omega(r)$, we similarly calculate Kretschmann invariant for the generalised solution \eqref{eq:CS-omega-soln}. Therefore, we obtain
\begin{equation}\label{eq:Ksq-omega}
     R^{\alpha \beta \gamma \delta} R_{\alpha \beta \gamma \delta} \simeq \frac{48 M^2}{r^6}+ \frac{192 M C_3^2}{r^3}\left(\frac{1}{3}+\cos 2\theta\right).
\end{equation}
Again the first term is associated with Schwarzschild solution, while the second term, $O(r^{-3})$, is associated to the rotation. In comparison to Kerr spacetime, Eq. \eqref{eq:Ksq-Kerr}, the frame dragging effect decays more slowly than the corresponding term in Kerr and than Schwarzschild term as well. On the contrary to the KMT case when $\omega\equiv \omega(r)$, the behaviour of Kretschmann invariant is asymptotically flat everywhere with no singularity along the rotational axis as in the KMT case, namely Eq. \eqref{eq:Ksq-omegar}. It is also expected to produce stronger dragging effect than the corresponding term in Eq. \eqref{eq:Ksq-omegar}. In the following section, we investigate possible consequences of the obtained solution on astrophysical scales.
%%%%%%%%%%%%%%%%%%%%%%%%%%%%%%%%%%%%%%%%%%%%%%%%%%%%%%%%%%%%%%%%%%%%%%%%%%%
\section{Rotation curve of Spiral Galaxies}
\label{sec:rotation-curve}
%%%%%%%%%%%%%%%%%%%%%%%%%%%%%%%%%%%%%%%%%%%%%%%%%%%%%%%%%%%%%%%%%%%%%%%%%%%
As a first step towards confronting the CS model, which is characterised by the spacetime metric \eqref{eq:slow-rot-sptime} with the CS solution \eqref{eq:CS-omega-soln}, with rotational curve observational data, we obtain the trajectories of a massive test particle with mass $m$ in the present spacetime configuration. These are determined by the geodesic equations
\begin{equation}
    \frac{d^2 x^\alpha}{d\tau^2}+\Gamma{^\alpha}{_{\mu\nu}}\frac{d x^\mu}{d\tau}\frac{d x^\nu}{d\tau}=0,
\end{equation}
where $\tau$ is the proper time characterises the above trajectory and $\Gamma$ is Levi-Civita linear connection. For the $\theta$-component, we set $x^\alpha=\theta$ which yields
\begin{equation}
   \frac{d^2 \theta}{d\tau^2}+\frac{2}{r}\frac{dr}{d\tau}\frac{d\theta}{d\tau}-\sin\theta \cos\theta \left[\left(\frac{d \phi}{d\tau}\right)^2-\omega(r,\theta)\frac{d\phi}{d\tau}\frac{dt}{d\tau}\right]=0.
\end{equation}
Taking initially the inclination angle $\theta=\theta_0=\pi/2$ and $d\theta_0/d\tau=0$, we have $d\theta/d\tau=0$. This shows that the motion of a test particle is planer motion in a good agreement with motion of particles on the galaxy disk. Without loss of generality, we restrict ourselves to circular motion of a test particle with mass $m$ in the edge-on disk\footnote{The stellar motion in the galactic disk can be, in practice, corrected due to different inclination angles as will be shown in {Section {\ref{sec:data_model}}}.}, i.e. inclination angle $\theta=\pi/2$, to confront the present model with the observed rotational velocity data of spiral galaxies. Since the spacetime metric does not depend on time nor azimuth angle, the components of the four momenta $p_t=-m E$ and $p_\phi=m L$ are conserved on a trajectory where $E$ and $L$ are energy and angular momentum per unit mass. Thus, it is easy to show that the first integral of Euler-Lagrange equation of $t$, $\phi$ and $r$ components can be written as
\begin{eqnarray}
    \frac{dt}{d\tau}&=&\frac{E-\omega L}{A(r)+ r^2 \omega(r)^2}, \label{eq:tdot}\\
    \frac{d\phi}{d\tau}&=&\frac{L}{r^2}+\omega(r)\frac{E-\omega L}{A(r)+ r^2 \omega(r)^2}, \label{eq:phidot}\\
    \frac{dr}{d\tau}&=&\sqrt{\left(1+\frac{L^2}{r^2}\right) A(r)+\frac{(E-\omega(r) L)^2 A(r)}{A(r)+ r^2 \omega(r)^2}}. \label{eq:rdot}
\end{eqnarray}
For the static limit $\omega\to 0$, the above equations reduce to Schwarzschild solution. For a circular orbit, using Equations \eqref{eq:tdot} and \eqref{eq:phidot}, we obtain the circular velocity
\begin{equation}\label{eq:circ_vel}
    v_{circ}=r\frac{d\phi}{dt}=\frac{A(r) L}{r(E-\omega(r) L)}+\frac{r \omega(r) E}{E-\omega(r) L},
\end{equation}
where $E$ and $L$ are determined using the circular motion constraints, i.e. $\frac{dr}{d\tau}=0=\frac{d^2 r}{d\tau^2}$. By virtue of Equation \eqref{eq:rdot}, we determine
\begin{eqnarray}
    E&=&\pm\frac{r-2M}{\sqrt{r(r-3M)}}+ \sqrt{\frac{r^2 M}{r-3M}}\left[\omega(r) + \frac{r(r-2M)}{2(r-3M)}\omega_{,r}\right]\nonumber +O(\epsilon^2), \\
    L&=&\pm r\sqrt{\frac{M}{r-3M}}+\frac{r^4(r-2M)}{2\sqrt{r(r-3M)^3}}\omega_{,r}+O(\epsilon^2).
\end{eqnarray}
\begin{figure}
    \centering
    \includegraphics[width=0.7\textwidth]{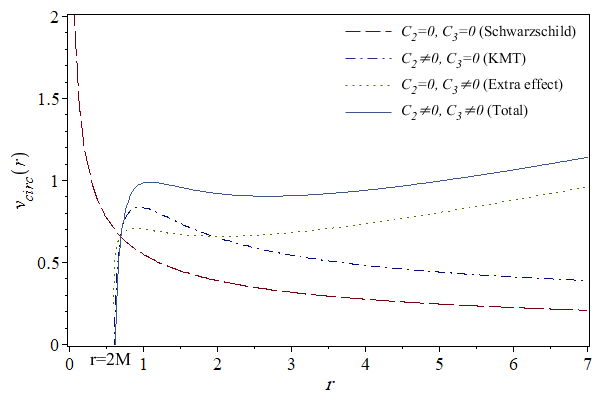}
    \caption{A schematic plot of the total rotational velocity curve \eqref{eq:circ_vel2x} associated with individual term contribution where the longdashed curve represents exactly Schwarzschild contribution ($C_2=0$ and $C_3=0$), i.e. $v_{circ}=\sqrt{M/r}$, the dashdotted curve represents KMT model ($C_2\neq0$, $C_3=0$), the dotted curve represents the velocity where $C_2=0$ and $C_3\neq 0$ and the solid curve represents the total velocity curve which gives rise to the present CS model.}
    \label{fig:schematic_plot}
\end{figure}
Substituting into Equation \eqref{eq:circ_vel}, where the frame-dragging effect in CS gravity is obtained by \eqref{eq:CS-omega-soln}, we give the circular velocity
\begin{eqnarray}\label{eq:circ_vel2x}
    v_{circ}&=&\sqrt{\frac{M}{r}}+\frac{C_2}{2r^2}\left[\right.r^2+4M^2+8M^2\ln(r-2M)+4rM\left.\right]\nonumber\\
            &+&\frac{C_3}{r^2}\Bigg\{12M^3-24M^3\ln(r-2M)-3r^2M-2r^312rM^2\Bigg\}.
\end{eqnarray}
For $M/r\ll 1$, at the distant outer regions of the rotational curve, following same approximations used in KMT model \citep{Konno:2008np}, the circular velocity reduces to
\begin{equation}\label{eq:circ_vel2}
    v_{circ}\approx \sqrt{\frac{M}{r}}+\frac{C_2}{2}+C_3(2r-7M).
\end{equation}
The above equation is the basic equation to calculate the rotational velocity of a massive particle in CSMG. Clearly, it reduces to KMT model where $C_3=0$. Indeed, the constant term associated to $C_2$ has been advocated to explain the flat part of a rotation curve with no need to dark matter \cite{Konno:2008np}. However, this claim has been never tested with observational data \cite{Alexander:2009tp}. The additional linear term in $r$ associated to the constant of integration $C_3$ could be potentially useful to describe galaxy rotation curves, while the first term in Equation \eqref{eq:circ_vel2} represents the Keplerian rotation curve (Schwarzschild solution). In Fig.~\ref{fig:schematic_plot}, we give a schematic plot showing individual contribution of each term  to the total velocity curve. We note that we exaggerate Schwarzschild radius $r=2M$, which determines the $r$-intercept of the curve, for better plotting purposes. The constant term $C_2$ alone seems to be not able to describe the outer region flat part of the curve unless it has a large value, which in turn pushes the curve peak to higher values not consistent with a reasonable mass choice. On the contrary, a reasonable choice of ``\textit{small}" value of the constant $C_2$ renders the velocity curve almost compatible with the Keplerian curve at the outer region. In the following section, we confront the obtained CS solution \eqref{eq:circ_vel2} with observational data in comparison to KMT model \cite{Konno:2008np}.
%%%%%%%%%%%%%%%%%%%%%%%%%%%%%%%%%%%%%%%%%%%%%%%%%%%%%%%%%%%%%%%
\section{Testing CSMG with UGC11455 galaxy rotation curve}
\label{sec:data_model}
%%%%%%%%%%%%%%%%%%%%%%%%%%%%%%%%%%%%%%%%%%%%%%%%%%%%%%%%%%%%%%%%
In this section, we give an example to constrain the parameter space of the KMT and the present models by fitting the rotation curve of UGC11455 galaxy as obtained by SPARC database \citep{Lelli:2016zqa}. The UGC11455 galaxy is characterised by a non-bulge profile, which simplifies the fitting procedure by reducing the parameter space (i.e. to include only the stellar disk component). The number of points are enough to ensure high resolution for the fitting purpose, where the effective radius $R_{\rm eff}=10.06$ kpc and scale radius $r_{s}=5.93$ kpc. This galaxy is classified as a High Surface Brightness (HSB), with effective surface brightness $\Sigma_{\rm eff}=571.26 L_\odot$pc$^{-2}$; this is to ensure that the galaxy is dominated by stellar matter rather than dark matter around the centre. Additionally, its disk inclination $i > 30^{\circ}$, which allows to ignore the bias due to galaxy's disk inclination. However, we account for the disk inclination by modifying the observed velocities by a factor of $1/\sin(i)$.

\subsection{Methodology}
For CSMG model, we mainly use the rotational velocity equation \eqref{eq:circ_vel2} giving up the geometric units assumption $G=c=1$. It proves convenient to write the free parameters of a model in dimensionless forms, therefore we use the transformation $C_2\to c C_2$ and $C_3\to H_0 C_3$ to obtain $C_2$ and $C_3$ dimensionless, where $H_0$ is Hubble constant with a dimension [T$^{-1}$]. We note that the CS solution \eqref{eq:CS-omega-soln} of the metric \eqref{eq:slow-rot-sptime} has been obtained for vacuum. However, galaxies have visible baryons matter with mass distributions, $M(r)$, up to large radial distances. Therefore, baryons matter with pressure $p(r)$ and density $\rho(r)$ profiles, i.e. equation of state ($p\equiv p(\rho)$), need to be imposed into the field equations similar to interior solution of stellar models where the metric potential, i.e.
\begin{equation}\label{eq:nonvsoln}
    g_{tt}=A(r)=1-\frac{2 G M(r)}{c^2 r},
\end{equation}
and the mass function
\begin{equation}
    M(r)=4\pi \int_{\xi=0}^r \rho(\xi) \xi^2 d\xi.
\end{equation}
However, in the present case, the relativistic effect of visible matter pressure in the galaxies is almost negligible. In this sense, the above solution, namely Equation \eqref{eq:nonvsoln}, is a good approximation to the full hydrostatic equilibrium configuration given by the TOV equation at weak field limit. Thus, we rewrite the rotational velocity \eqref{eq:circ_vel2} up to $O(r^{-1})$ as follows
\begin{equation}\label{eq:circ_vel3}
    v_{circ}\approx \sqrt{\frac{G M(r)}{r}}+\frac{c}{2}C_2 + 2 r H_0 C_3  \left(1-\frac{7}{2}\frac{G M(r)}{c^2 r}\right),
\end{equation}
where the mass distribution of the baryons visible matter for every galaxy is the following parameterised form \citep{Lelli:2016zqa}
\begin{equation}\label{eq:mass_dist}
    \frac{G M(r)}{r}=V_\text{bar}^2= \Upsilon_{\rm bulge} V_{\rm bulge}^2 + \Upsilon_{\rm disk} V_{\rm disk}^2 + V_{\rm gas}|V_{\rm gas}|,
\end{equation}
with $\Upsilon_{\rm bulge,disk}$ being two free parameters (mass-to-light ratios) associated to the contributions of the bulge velocity, $V_{\rm bulge}$, and the disk velocity, $V_{\rm disk}$, to the baryon velocity. This is in addition to the gas contribution $V_{\rm gas}$, where absolute value is used since the gas near the centre may have opposite rotational direction. We remind that no bulge is assumed in the present study to simplify the fitting procedure. The above formula is appropriate with static spherically symmetric solution. However, it is still useful in our procedure, since the Keplerian term in the circular velocity \eqref{eq:circ_vel3} is dominant. Plugging \eqref{eq:mass_dist} into \eqref{eq:circ_vel3}, we finally write the CS rotational velocity
\begin{equation}\label{eq:CS_rot_vel}
    V_{\rm CS}:= v_{circ} = V_{\rm bar} + \frac{c}{2}C_2 + 2 r H_0 C_3 \left(1 -\frac{7}{2} \frac{V^2_{\rm bar}}{c^2}\right),
\end{equation}
where the frame dragging effect introduces $C_2$ and $C_3$ as free parameters of the CS model to be fit with observational data.

We use Markov Chain Monte Carlo (MCMC) method to fit the observed rotation curves of UGC11455 galaxy. Additionally, we use the emcee python package \citep{2013PASP..125..306F} to map the posterior distribution of the fitting parameters. The priors are taken such that lognormal probability distribution function (PDF) is imposed on $\Upsilon_{\rm disk}$ around fiducial value 0.5 $M_\odot/L_\odot$ with standard deviation of 0.1 dex as suggested by stellar population models \citep{2017ApJ...836..152L,2019MNRAS.483.1496S}. The priors of the tested models, $\vec{\alpha}=\{\alpha_i\}$, are flat priors, where $C_2 \in [0, 10]$ and $C_3 \in [-140, 420]$. For KMT model, we set $C_3=0$. We use the likelihood function,
\begin{equation}
    {\cal L}(\chi)  =  \exp(-\frac{1}{2}\chi^2),
\end{equation}
where $\chi^2$ is defined as
\begin{equation}
    \chi^2 = \sum_{i=1}^{N} \left(\frac{V_{\rm obs}(R_i)-V_{\rm CS}(\Upsilon_{\rm disk}, \vec{\alpha},R_i)}{\sigma_{V_{\rm obs}(R_i)}}\right)^2,
\end{equation}
and $V_{\rm obs}$ is the $N$th observed rotational velocity of the UGC11455 galaxy and $\sigma_V$ is the corresponding uncertainty of $V_{\rm obs}$ as provided by SPARC database \citep{Lelli:2016zqa}. The MCMC chains are initialised with 164 random walkers using the standard emcee affine-invariant sampler. The number of steps are set to 1000 during the burn-in period for each walker. Then, we reset the sampler and run longer chains with number of steps equal to 4000 to ensure chains convergence.

\subsection{Results}\label{sec:results}
\begin{figure}[t]
    \centering
    \includegraphics[width=0.5\linewidth]{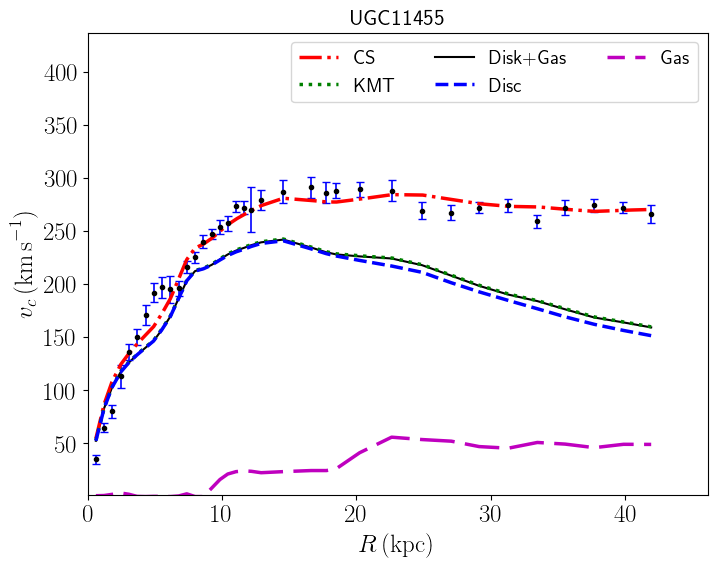}
    \caption{The rotational velocity curve of the galaxy UGC11455. Obviously the KMT model cannot fit the data at large distances which are supposed to be subjected to the modified gravity. This leads to the conclusion that the KMT model cannot fit galaxy rotation curves in practice. The CSMG fits the observed curve well where the outer region is dominated by the $C_3$-term in Equation \eqref{eq:CS_rot_vel}.}
    \label{fig:KMT1}
\end{figure}

In this subsection, we mainly present the obtained results by applying the methodology mentioned above to constrain the parameter space of the KMT and CSMG models. Nevertheless, we also explain the failure of the KMT model to fit real rotation curves data as previously discussed by the schematic plot in Fig. \ref{fig:schematic_plot}. We do this by setting $C_3=0$ in the circular velocity given by Eq. \eqref{eq:CS_rot_vel} to constrain the free parameter $C_2$. Then, we apply the above mentioned procedure to UGC11455 galaxy rotation curve, and consequently we obtain the velocity curve as shown by Fig. \ref{fig:KMT1}. In this case, the best fit of the parameter $C_2=0.06746$ ($\sim 2.02$ km/s), which is too small to maintain the flat part ($250-300$ km/s) of the rotation curve. On another word, the KMT model (dotted curve) is practically reproduces the Keplerian velocity curve. On the other hand, if one requires $C_2$ to be larger in order to fit the flatten region of the curve, it would much shift the curve upward spoiling up the inner region $R<10$ kpc by requiring large mass-to-light ratio $\Upsilon_{\rm disk}$, which results in nonphysical masses above reasonable values.
\begin{table}[t]
    \centering
    \caption{A comparison between KMT solution and the present study.}
    \begin{tabular}{|c|c|c|}
\hline
     & KMT & present solution\\
     \hline
frame-dragging     & Equation \eqref{eq:CS-omegar-soln}& Equation \eqref{eq:CS-omega-soln}\\[5pt]
     \hline
At $r\gg M$     & $g_{t\phi}=C_1+C_2 r$ & $g_{t\phi}=C_1+C_2 r+2C_3 r^2 \sin^2 \theta$\\[5pt]
     \hline
Cotton tensor     & vanishes & does not vanish\\[5pt]
     %\hline
comment     &GR solution& CS contributes non-trivially\\[5pt]
\hline
Kretschmann scalar     & $\sim \frac{48 M^2}{r^6}-\frac{4 C_2^2}{r^4 \sin^4 \theta}$ & $\sim \frac{48 M^2}{r^6}+\frac{192 M C_3^2}{r^3}(1/3+\cos 2\theta)$\\[5pt]
     %\hline
comment     &diverges at $\theta=0$ and $\pi$ & asymptotically flat everywhere\\[5pt]
\hline
circular velocity     & $\sim \sqrt{M/r}+C_2/2$ & $\sim \sqrt{M/r}+C_2/2+C_3(2r-7M)$\\[5pt]
     \hline
rotation curve     & does not fit & fits well where the perturbation\\
                   &              & term dominates, which violates\\
                   &              & the model assumption.\\[5pt]
\hline
\end{tabular}
        \label{tab:comparison}
\end{table}

Next, we turn our attention to the CSMG solution, namely Equation \eqref{eq:CS_rot_vel}.  For the same rotation curve of UGC11455 galaxy, the best fit values of the free parameters are $\Upsilon_{\rm disk}=0.543\pm 0.011$ $M_\odot/L_\odot$ in agreement with the predicted value as suggested by stellar population models, while $10^{-4} C_2=0.0347^{+0.035}_{-0.0056}$ and $C_3=18.30\pm 0.62$ with $\chi^2_\nu:=\chi^2/{\rm dof}=3.82329$, where dof is the degrees of freedom of the model. We note that the $C_3$-term is associated with the radial distance. On the contrary to the KMT, we find that the CSMG model (dashdotted curve) fits the data well as shown by Fig. \ref{fig:KMT1}, since the $C_3$-term contribution to the outer region of the velocity curve is considerable and has the dominant effect. However, this violates the model assumption, where the perturbation around Schwarzschild BH should be small with respect to the leading term. We summarise the results of the present study in Table \ref{tab:comparison}, which traces the main features of KMT solution and the present one.  
%%%%%%%%%%%%%%%%%%%%%%%%%%%%%%%%%%%%%%%%%%%%%%%%%%
\section{Summary and conclusion}
\label{sec:Summary}
%%%%%%%%%%%%%%%%%%%%%%%%%%%%%%%%%%%%%%%%%%%%%%%%%%
In the present work, we revisited slow rotating BHs solution in CSMG using the perturbation expansion around the Schwarzschild. In particular, the nondynamical CSMG with noncanonical CS scalar, which has been proposed by \cite{Konno:2007ze} to explain the flatness of rotational velocity curves of spiral galaxies \citep{Konno:2008np}. The present study showed that the KMT model assumption, which requires the frame dragging to have a radial dependence only, derives the Cotton tensor to vanish up to first order. In effect, the CSMG and GR solutions are identical, and then CSMG does not provide any new insights. However, by dropping the KMT assumption, we showed that the full frame dragging should have radial as well as zenith angle dependence.

Unlike KMT model, the present CS solution differs from GR and reduces to KMT as a particular case. Indeed, the CS solution fulfils GR field equations, since $G_{\mu\nu}=0$ is satisfied. We note that the opposite conclusion is not true. On another word, the GR solution does not necessarily satisfy the divergence-free Cotton tensor in CSMG, which is the case in the present study. We showed that the tidal gravitational force in terms of Kretschmann scalar, $K$, of the new solution has an additional term, $\propto 1/r^3$, associated to the leading term of Schwarzschild solution, i.e.  $K=48M^2/r^6$. Notably, the present solution provides an asymptotically flat Kretschmann invariant everywhere and does not diverge along any specific direction. This is unlike KMT, in which Kretschmann invariant diverges along $\theta=0$ and $\theta =\pi$ due to presence of $1/(r\sin \theta)^{4}$ term. In addition, the present CS solution predicts frame dragging effect, which decays more slowly than the corresponding term in Kerr, $\propto 1/r^8$, and than Schwarzschild term, $\propto 1/r^6$, as well. In this sense, the new solution should modify the static spherically symmetric solution at large distances, and then it worths to be investigated via astrophysical observations of rotational velocities on the galactic scales.

We confronted the KMT and CSMG models with the rotational velocity curve of UGC1145 galaxy using MCMC method to fit the parameter space of each model. The UGC1145 galaxy is characterised by HSB, no bulge, more than seven points and inclination $i > 30^{\circ}$ to ensure that the galaxies are dominated by stellar matter rather than dark matter around the centre with good statistical resolution and no bias due to galaxy's disk inclination. We concluded that the KMT model reproduces the Keplerian velocity, since $C_2$ term contributes to the velocity by a small fraction $\sim 2$ km/s, and consequently it cannot fit the rotation curve. On the other hand, the present solution fits the data well with reasonable $\chi^2_\nu$, where the contribution of $C_3$ becomes dominant at the outer region of the velocity curve. However, this contribution violates the model assumption by assuming the correction to Schwarzschild solution to be small with respect to the leading term, i.e. the Keplerian velocity. This conclusion should be valid to any perturbative solution around Schwarzschild BH, which should be small. Therefore, it would fail to explain the flatten part of the velocity curve at large distances from the galactic centre.

In conclusion, we may say that CS gravity is running out of options to explain galactic rotation curves. However, we may suggest two different approaches to the problem: (i) In presence of dark matter, CS modified gravity may require different amount of dark matter than the standard $\Lambda$CDM scenario, which could be tested on cosmological scales with clustering rates of galaxies. Similar treatments in generic framework of modified gravity can be found in \cite{deAlmeida:2018kwq}. On the contrary, if the analysis requires $C_2$ and $C_3$ to be null, this may provide a strict constraint on deviations from Schwarzschild BH solution even with noncanonical choices of the CS scalar within CSMG. This is in agreement with Ref. \cite{Grumiller:2007rv}, where the standard choice of the CS scalar is applied. (ii) It has been argued that \textit{physical} spinning BH solution may exist in CSMG, if energy-momentum conservation is broken \cite{Grumiller:2007rv}.  Then, we may suggest another approach by considering matter-CS nonminimal coupling, which should induce extra force orthogonal to the four-velocity rendering the motion of test particles non-geodesic \cite{Harko:2014gwa}. In general, the Newtonian limit of such scenario obtains an acceleration law similar to MOND. We leave this task for a future study.
\bibliographystyle{IEEEtran}
\bibliography{csmg_v3}

\begin{thebibliography}{10}
\providecommand{\url}[1]{#1}
\csname url@samestyle\endcsname
\providecommand{\newblock}{\relax}
\providecommand{\bibinfo}[2]{#2}
\providecommand{\BIBentrySTDinterwordspacing}{\spaceskip=0pt\relax}
\providecommand{\BIBentryALTinterwordstretchfactor}{4}
\providecommand{\BIBentryALTinterwordspacing}{\spaceskip=\fontdimen2\font plus
\BIBentryALTinterwordstretchfactor\fontdimen3\font minus
  \fontdimen4\font\relax}
\providecommand{\BIBforeignlanguage}[2]{{%
\expandafter\ifx\csname l@#1\endcsname\relax
\typeout{** WARNING: IEEEtran.bst: No hyphenation pattern has been}%
\typeout{** loaded for the language `#1'. Using the pattern for}%
\typeout{** the default language instead.}%
\else
\language=\csname l@#1\endcsname
\fi
#2}}
\providecommand{\BIBdecl}{\relax}
\BIBdecl

\bibitem{Konno:2007ze}
K.~Konno, T.~Matsuyama, and S.~Tanda, ``{Does a black hole rotate in
  Chern-Simons modified gravity?}'' \emph{Phys. Rev. D}, vol.~76, p. 024009,
  2007.

\bibitem{RevModPhys.19.361}
\BIBentryALTinterwordspacing
G.~M. Clemence, ``The relativity effect in planetary motions,'' \emph{Rev. Mod.
  Phys.}, vol.~19, pp. 361--364, Oct 1947. [Online]. Available:
  \url{https://link.aps.org/doi/10.1103/RevModPhys.19.361}
\BIBentrySTDinterwordspacing

\bibitem{doi:10.1098/rsta.1920.0009}
\BIBentryALTinterwordspacing
F.~W. Dyson, A.~S. Eddington, and C.~Davidson, ``Ix. a determination of the
  deflection of light by the sun's gravitational field, from observations made
  at the total eclipse of may 29, 1919,'' \emph{Philosophical Transactions of
  the Royal Society of London. Series A, Containing Papers of a Mathematical or
  Physical Character}, vol. 220, no. 571-581, pp. 291--333, 1920. [Online].
  Available:
  \url{https://royalsocietypublishing.org/doi/abs/10.1098/rsta.1920.0009}
\BIBentrySTDinterwordspacing

\bibitem{1981SciAm.245d..74W}
J.~M. {Weisberg}, J.~H. {Taylor}, and L.~A. {Fowler}, ``{Gravitational waves
  from an orbiting pulsar},'' \emph{Scientific American}, vol. 245, pp. 74--82,
  Oct. 1981.

\bibitem{2010ApJ...722.1030W}
J.~M. {Weisberg}, D.~J. {Nice}, and J.~H. {Taylor}, ``{Timing Measurements of
  the Relativistic Binary Pulsar PSR B1913+16},'' \emph{Astrophys. J.}, vol.
  722, no.~2, pp. 1030--1034, Oct. 2010.

\bibitem{Will:2014kxa}
C.~M. Will, ``{The Confrontation between General Relativity and Experiment},''
  \emph{Living Rev. Rel.}, vol.~17, p.~4, 2014.

\bibitem{LIGOScientific:2016aoc}
B.~P. Abbott \emph{et~al.}, ``{Observation of Gravitational Waves from a Binary
  Black Hole Merger},'' \emph{Phys. Rev. Lett.}, vol. 116, no.~6, p. 061102,
  2016.

\bibitem{LIGOScientific:2017ycc}
------, ``{GW170814: A Three-Detector Observation of Gravitational Waves from a
  Binary Black Hole Coalescence},'' \emph{Phys. Rev. Lett.}, vol. 119, no.~14,
  p. 141101, 2017.

\bibitem{LIGOScientific:2017vwq}
------, ``{GW170817: Observation of Gravitational Waves from a Binary Neutron
  Star Inspiral},'' \emph{Phys. Rev. Lett.}, vol. 119, no.~16, p. 161101, 2017.

\bibitem{EventHorizonTelescope:2019dse}
K.~Akiyama \emph{et~al.}, ``{First M87 Event Horizon Telescope Results. I. The
  Shadow of the Supermassive Black Hole},'' \emph{Astrophys. J. Lett.}, vol.
  875, p.~L1, 2019.

\bibitem{1970ApJ...159..379R}
V.~C. {Rubin} and J.~{Ford}, W.~Kent, ``{Rotation of the Andromeda Nebula from
  a Spectroscopic Survey of Emission Regions},'' \emph{Astrophys. J.}, vol.
  159, p. 379, Feb. 1970.

\bibitem{1980ApJ...238..471R}
V.~C. {Rubin}, J.~{Ford}, W.~K., and N.~{Thonnard}, ``{Rotational properties of
  21 SC galaxies with a large range of luminosities and radii, from NGC 4605
  (R=4kpc) to UGC 2885 (R=122kpc).}'' \emph{Astrophys. J.}, vol. 238, pp.
  471--487, Jun. 1980.

\bibitem{1991MNRAS.249..523B}
K.~G. {Begeman}, A.~H. {Broeils}, and R.~H. {Sanders}, ``{Extended rotation
  curves of spiral galaxies : dark haloes and modified dynamics.}'' \emph{Mon.
  Not. Roy. Astron. Soc.}, vol. 249, p. 523, Apr. 1991.

\bibitem{Corbelli:1999af}
E.~Corbelli and P.~Salucci, ``{The Extended Rotation Curve and the Dark Matter
  Halo of M33},'' \emph{Mon. Not. Roy. Astron. Soc.}, vol. 311, pp. 441--447,
  2000.

\bibitem{Planck:2018vyg}
N.~Aghanim \emph{et~al.}, ``{Planck 2018 results. VI. Cosmological
  parameters},'' \emph{Astron. Astrophys.}, vol. 641, p.~A6, 2020, [Erratum:
  Astron.Astrophys. 652, C4 (2021)].

\bibitem{Jungman:1995df}
G.~Jungman, M.~Kamionkowski, and K.~Griest, ``{Supersymmetric dark matter},''
  \emph{Phys. Rept.}, vol. 267, pp. 195--373, 1996.

\bibitem{Craig:2013cxa}
N.~Craig, ``{The State of Supersymmetry after Run I of the LHC},'' in
  \emph{{Beyond the Standard Model after the first run of the LHC}}, 9 2013.

\bibitem{COSINE-100:2021xqn}
G.~Adhikari \emph{et~al.}, ``{Strong constraints from COSINE-100 on the DAMA
  dark matter results using the same sodium iodide target},'' \emph{Sci. Adv.},
  vol.~7, no.~46, p. abk2699, 2021.

\bibitem{1983ApJ...270..365M}
M.~{Milgrom}, ``{A modification of the Newtonian dynamics as a possible
  alternative to the hidden mass hypothesis.}'' \emph{Astrophys. J.}, vol. 270,
  pp. 365--370, Jul. 1983.

\bibitem{1983ApJ...270..371M}
------, ``{A modification of the Newtonian dynamics - Implications for
  galaxies.}'' \emph{Astrophys. J.}, vol. 270, pp. 371--383, Jul. 1983.

\bibitem{Skordis:2020eui}
C.~Skordis and T.~Zlosnik, ``{New Relativistic Theory for Modified Newtonian
  Dynamics},'' \emph{Phys. Rev. Lett.}, vol. 127, no.~16, p. 161302, 2021.

\bibitem{Chae:2022oft}
K.-H. Chae, ``{Distinguishing Dark Matter, Modified Gravity, and Modified
  Inertia with the Inner and Outer Parts of Galactic Rotation Curves},''
  \emph{Astrophys. J.}, vol. 941, no.~1, p.~55, 2022.

\bibitem{moffat2004modified}
J.~W. Moffat, ``Modified gravitational theory and galaxy rotation curves,''
  2004.

\bibitem{Bertolami:2007gv}
O.~Bertolami, C.~G. Boehmer, T.~Harko, and F.~S.~N. Lobo, ``{Extra force in
  f(R) modified theories of gravity},'' \emph{Phys. Rev. D}, vol.~75, p.
  104016, 2007.

\bibitem{Harko:2011kv}
T.~Harko, F.~S.~N. Lobo, S.~Nojiri, and S.~D. Odintsov, ``{$f(R,T)$ gravity},''
  \emph{Phys. Rev. D}, vol.~84, p. 024020, 2011.

\bibitem{Moffat:2013sja}
J.~W. Moffat and S.~Rahvar, ``{The MOG weak field approximation and
  observational test of galaxy rotation curves},'' \emph{Mon. Not. Roy. Astron.
  Soc.}, vol. 436, pp. 1439--1451, 2013.

\bibitem{Capozziello:2013yha}
S.~Capozziello, T.~Harko, T.~S. Koivisto, F.~S.~N. Lobo, and G.~J. Olmo,
  ``{Galactic rotation curves in hybrid metric-Palatini gravity},''
  \emph{Astropart. Phys.}, vol. 50-52, pp. 65--75, 2013.

\bibitem{Moffat:2014pia}
J.~W. Moffat and V.~T. Toth, ``{Rotational velocity curves in the Milky Way as
  a test of modified gravity},'' \emph{Phys. Rev. D}, vol.~91, no.~4, p.
  043004, 2015.

\bibitem{Lelli:2016zqa}
F.~Lelli, S.~S. McGaugh, and J.~M. Schombert, ``{SPARC: Mass Models for 175
  Disk Galaxies with Spitzer Photometry and Accurate Rotation Curves},''
  \emph{Astron. J.}, vol. 152, p. 157, 2016.

\bibitem{Panpanich:2018cxo}
S.~Panpanich and P.~Burikham, ``{Fitting rotation curves of galaxies by de
  Rham-Gabadadze-Tolley massive gravity},'' \emph{Phys. Rev. D}, vol.~98,
  no.~6, p. 064008, 2018.

\bibitem{Chae:2020omu}
K.-H. Chae, F.~Lelli, H.~Desmond, S.~S. McGaugh, P.~Li, and J.~M. Schombert,
  ``{Testing the Strong Equivalence Principle: Detection of the External Field
  Effect in Rotationally Supported Galaxies},'' \emph{Astrophys. J.}, vol. 904,
  no.~1, p.~51, 2020, [Erratum: Astrophys.J. 910, 81 (2021)].

\bibitem{Burikham:2023bil}
P.~Burikham, T.~Harko, K.~Pimsamarn, and S.~Shahidi, ``{Dark matter as a Weyl
  geometric effect},'' \emph{Phys. Rev. D}, vol. 107, no.~6, p. 064008, 2023.

\bibitem{Clowe:2003tk}
D.~Clowe, A.~Gonzalez, and M.~Markevitch, ``{Weak lensing mass reconstruction
  of the interacting cluster 1E0657-558: Direct evidence for the existence of
  dark matter},'' \emph{Astrophys. J.}, vol. 604, pp. 596--603, 2004.

\bibitem{Markevitch:2003at}
M.~Markevitch, A.~H. Gonzalez, D.~Clowe, A.~Vikhlinin, L.~David, W.~Forman,
  C.~Jones, S.~Murray, and W.~Tucker, ``{Direct constraints on the dark matter
  self-interaction cross-section from the merging galaxy cluster 1E0657-56},''
  \emph{Astrophys. J.}, vol. 606, pp. 819--824, 2004.

\bibitem{Pardo:2020epc}
K.~Pardo and D.~N. Spergel, ``{What is the price of abandoning dark matter?
  Cosmological constraints on alternative gravity theories},'' \emph{Phys. Rev.
  Lett.}, vol. 125, no.~21, p. 211101, 2020.

\bibitem{Carroll:1989vb}
S.~M. Carroll, G.~B. Field, and R.~Jackiw, ``{Limits on a Lorentz and Parity
  Violating Modification of Electrodynamics},'' \emph{Phys. Rev. D}, vol.~41,
  p. 1231, 1990.

\bibitem{Jackiw:2003pm}
R.~Jackiw and S.~Y. Pi, ``{Chern-Simons modification of general relativity},''
  \emph{Phys. Rev. D}, vol.~68, p. 104012, 2003.

\bibitem{ALVAREZGAUME1984269}
\BIBentryALTinterwordspacing
L.~Alvarez-Gaum\'{e} and E.~Witten, ``Gravitational anomalies,'' \emph{Nuclear
  Physics B}, vol. 234, no.~2, pp. 269--330, 1984. [Online]. Available:
  \url{https://www.sciencedirect.com/science/article/pii/055032138490066X}
\BIBentrySTDinterwordspacing

\bibitem{2012ssti.book.....G}
M.~B. {Green}, J.~H. {Schwarz}, and E.~{Witten}, \emph{{Superstring Theory,
  Volume 1: Introduction}}, 2012.

\bibitem{1989IJMPA...4.1493A}
A.~{Ashtekar}, A.~P. {Balachandran}, and S.~{Jo}, ``{The CP Problem in Quantum
  Gravity},'' \emph{International Journal of Modern Physics A}, vol.~4, no.~6,
  pp. 1493--1514, Jan. 1989.

\bibitem{Alexander:2009tp}
S.~Alexander and N.~Yunes, ``{Chern-Simons Modified General Relativity},''
  \emph{Phys. Rept.}, vol. 480, pp. 1--55, 2009.

\bibitem{Grumiller:2007rv}
D.~Grumiller and N.~Yunes, ``{How do Black Holes Spin in Chern-Simons Modified
  Gravity?}'' \emph{Phys. Rev. D}, vol.~77, p. 044015, 2008.

\bibitem{Konno:2008np}
K.~Konno, T.~Matsuyama, Y.~Asano, and S.~Tanda, ``{Flat rotation curves in
  Chern-Simons modified gravity},'' \emph{Phys. Rev. D}, vol.~78, p. 024037,
  2008.

\bibitem{Yunes:2009hc}
N.~Yunes and F.~Pretorius, ``{Dynamical Chern-Simons Modified Gravity. I.
  Spinning Black Holes in the Slow-Rotation Approximation},'' \emph{Phys. Rev.
  D}, vol.~79, p. 084043, 2009.

\bibitem{Lue:1998mq}
A.~Lue, L.-M. Wang, and M.~Kamionkowski, ``{Cosmological signature of new
  parity violating interactions},'' \emph{Phys. Rev. Lett.}, vol.~83, pp.
  1506--1509, 1999.

\bibitem{Alexander:2006mt}
S.~H.~S. Alexander, ``{Is cosmic parity violation responsible for the anomalies
  in the WMAP data?}'' \emph{Phys. Lett. B}, vol. 660, pp. 444--448, 2008.

\bibitem{2013PASP..125..306F}
D.~{Foreman-Mackey}, D.~W. {Hogg}, D.~{Lang}, and J.~{Goodman}, ``{emcee: The
  MCMC Hammer},'' \emph{PASP}, vol. 125, no. 925, p. 306, Mar. 2013.

\bibitem{2017ApJ...836..152L}
F.~{Lelli}, S.~S. {McGaugh}, J.~M. {Schombert}, and M.~S. {Pawlowski}, ``{One
  Law to Rule Them All: The Radial Acceleration Relation of Galaxies},''
  \emph{Astrophys. J.}, vol. 836, no.~2, p. 152, Feb. 2017.

\bibitem{2019MNRAS.483.1496S}
J.~{Schombert}, S.~{McGaugh}, and F.~{Lelli}, ``{The mass-to-light ratios and
  the star formation histories of disc galaxies},'' \emph{Mon. Not. Roy.
  Astron. Soc.}, vol. 483, no.~2, pp. 1496--1512, Feb. 2019.

\bibitem{deAlmeida:2018kwq}
A.~O.~F. de~Almeida, L.~Amendola, and V.~Niro, ``{Galaxy rotation curves in
  modified gravity models},'' \emph{JCAP}, vol.~08, p. 012, 2018.

\bibitem{Harko:2014gwa}
T.~Harko and F.~S.~N. Lobo, ``{Generalized curvature-matter couplings in
  modified gravity},'' \emph{Galaxies}, vol.~2, no.~3, pp. 410--465, 2014.

\end{thebibliography}

\end{document}